\documentclass[iop]{emulateapj}

\usepackage{amsmath}
\usepackage{float}
\usepackage{booktabs}
\usepackage{hyperref}

\usepackage{color}

\begin{document}

\title{Future prospects for ground-based gravitational wave detectors --- The Galactic double neutron star merger rate revisited}

\author{Nihan Pol}
\author{Maura McLaughlin}
\author{Duncan R. Lorimer}
\affil{Department of Physics and Astronomy, West Virginia University, Morgantown, WV 26506-6315}
\affil{Center for Gravitational Waves and Cosmology, West Virginia University, Chestnut Ridge Research Building, Morgantown, West Virginia 26505}

\begin{abstract}
    We present the Galactic merger rate for double neutron star (DNS) binaries using the observed sample of eight DNS systems merging within a Hubble time. This sample includes the recently discovered, highly relativistic DNS systems J1757$-$1854 and J1946+2052, and is approximately three times the sample size used in previous estimates of the Galactic merger rate by Kim et al. Using this sample, we calculate the vertical scale height for DNS systems in the Galaxy to be $z_0 = 0.4 \pm 0.1$~kpc. We calculate a Galactic DNS merger rate of $\mathcal{R}_{\rm MW} = 42^{+30}_{-14}$~Myr$^{-1}$ at the 90\% confidence level. The corresponding DNS merger detection rate for Advanced LIGO is $\mathcal{R}_{\rm LIGO} = 0.18^{+0.13}_{-0.06} \times \left( D_{\rm r}/100 \ \rm Mpc \right)^3 \rm yr^{-1}$, where $D_{\rm r}$ is the range distance. Using this merger detection rate and the predicted range distance of 120--170~Mpc for the third observing run of LIGO (Laser Interferometer Gravitational-wave Observatory, Abbott et al., 2018), we predict, accounting for 90\% confidence intervals, that LIGO--Virgo will detect anywhere between zero and two DNS mergers. We explore the effects of the underlying pulsar population properties on the merger rate and compare our merger detection rate with those estimated using different formation and evolutionary scenario of DNS systems. As we demonstrate, reconciling the rates are sensitive to assumptions about the DNS population, including its radio pulsar luminosity function. Future constraints from further gravitational wave DNS detections and pulsar surveys anticipated in the near future should permit tighter constraints on these assumptions. 
\end{abstract}


\section{Introduction}      \label{intro}
    
    The first close binary with two neutron stars (NSs) discovered was PSR~B1913+16 \citep{1913_ht_discovery}. This double neutron star (DNS) system, known as the Hulse-Taylor binary, provided the first evidence for the existence of gravitational waves through measurement of orbital period decay in the system \citep{1913_gw_emission}. This discovery resulted in a Nobel Prize being awarded to Hulse and Taylor in 1993. The discovery of the Hulse-Taylor binary opened up exciting possibilities of studying relativistic astrophysical phenomena and testing the general theory of relativity and alternative theories of gravity in similar DNS systems \citep{stairs_dns_review}.
    
    Despite the scientific bounty on offer, relatively few DNS systems have been discovered since the Hulse-Taylor binary, with only 15 more systems discovered since. DNS systems are intrinsically rare since they require the binary system to remain intact with both components of the system undergoing supernova explosions to reach the final neutron star stage of their evolution. In addition, DNS systems are very hard to detect because of the large accelerations experienced by the two neutron stars in the system, which results in large Doppler shifts in their observed rotational periods \citep{Bagchi_gamma}.
    
    As demonstrated in the Hulse-Taylor binary, the orbit of these DNS systems decays through the emission of gravitational waves which eventually leads to the merger of the two neutron stars in the system \citep{1913_gw_emission}. DNS mergers are sources of gravitational waves that can be detected by ground-based detectors such as the Laser Interferometer Gravitational-Wave Observatory \citep[LIGO,][]{LIGO_detector_ref} in the USA and the Virgo detector \citep[][]{VIRGO_detector_ref} in Europe. Very recently, one such double neutron star (DNS) merger was observed by the LIGO-Virgo network
    \citep[][]{THE_DNS_merger} which was also detected across the electromagnetic spectrum \citep[][]{THE_DNS_merger_EM_assoc}, heralding a new age of multi-messenger gravitational wave astrophysics.
    
    We can predict the number of such DNS mergers that the LIGO-Virgo network will be able to observe by determining the merger rate in the Milky Way, and then extrapolate it to the observable volume of the LIGO-Virgo network. The first such estimates were provided by \citet[][]{Phinney_blue_lum_scaling} and \citet[][]{Narayan_1991} based on the DNS systems B1913+16 \citep{1913_ht_discovery} and B1534+12 \citep{1534_disc}. A more robust approach for calculating the merger rate was developed by \citet[][hereafter KKL03]{dunc_merger_1}, on the basis of which \citet{0737A_disc} and \citet[][]{A_effect_on_merger_rate, Kim_B_merger} were able to update the merger rate by including the Double Pulsar J0737--3039 system \citep{Lyne_dpsr, 0737A_disc}. 
    
    In the method described in KKL03, which we adopt in this work, we simulate the population of DNS systems like the ones we have detected by modelling the selection effects introduced by the different pulsar surveys in which these DNS systems are discovered or re-detected. This population of the DNS systems is then suitably scaled to account for the lifetime of the DNS systems and the number of such systems in which the pulsar beam does not cross our line of sight. We are only interested in those DNS systems that will merge within a Hubble time. Using this methodology, \citet[][]{Kim_B_merger} estimated the Galactic merger rate to be $\mathcal{R}_{\rm g} = 21^{+28}_{-14}$~Myr$^{-1}$ and the total merger detection rate for LIGO to be $\mathcal{R}_{\rm LIGO} = 8^{+10}_{-5}$~yr$^{-1}$, with errors quoted at the 95\% confidence interval, assuming a horizon distance of $D_{\rm h} = 445$~Mpc, and with the B1913+16 and J0737--3039 systems being the largest contributors to the rates.
    
    However, \citet[][]{use_range_not_horizon} recommend that the range distance, $D_{\rm r}$, is the correct distance estimate to convert from a merger rate density to a detection rate. For Euclidean geometry, this range distance is a factor of 2.264 smaller than the horizon distance, $D_{\rm h}$, which is the distance estimate used in all previous estimates of the LIGO detection rate. \citet[][]{LIGO_horizon_dist} calculate the range distance for the first and second observing runs of LIGO to be 60--80~Mpc and 60--100~Mpc respectively. The range distance for the upcoming third LIGO observing run is predicted to be in the range 120--170~Mpc \citep[][]{LIGO_horizon_dist}. Correspondingly, we correct the \citet[][]{Kim_B_merger} merger rate for LIGO by scaling their Galactic merger rate to a range distance of 100~Mpc \citep[Eq.~14 in][]{Kim_B_merger}. This gives us a revised detection rate for LIGO of
    \begin{equation}
            \displaystyle \mathcal{R}_{\rm LIGO} = 0.09^{+0.12}_{-0.06} \times \left( \frac{D_{\rm r}}{100 \, \rm Mpc} \right)^3 \rm yr^{-1}.
            \label{kim_rate}
        \end{equation}
    
    In this work, we include five new DNS systems into the estimation of the merger rate. Out of these five systems, two, J1757--1854 \citep{1757_disc}, with a time to merger of 76~Myr, and 1946+2052 \citep{1946_disc}, with a time to merger of 46~Myr, are highly relativistic systems that will merge on a timescale shorter than that of the Double Pulsar, which had the previous shortest time to merger of 85~Myr. The other DNS systems that we include in our analysis, J1906+0746 \citep{1906_disc}, J1756--2251 \citep{1756_disc}, and J1913+1102 \citep{1102_disc}, are not as relativistic, but are important to accurately modelling the complete Milky Way merger rate. These systems were not included in the previous studies due to insufficient evidence for them being DNS systems. However, \citet[][for J1906+0746]{1906_dns_evidence}, \citet[][for J1756--2251]{1756_dns_evidence}, and \citet[][for J1913+1102]{1102_dns_evidence} have established through timing observations that these are DNS systems.
    
    We tabulate the properties of all the known DNS binaries in the Milky Way, sorted by their time to merger, in Table~\ref{psr_table}. With the inclusion of the five additional systems, our sample size for calculating the merger rate is almost three times the one used in \citet[][]{Kim_B_merger}. In Section~\ref{survey_sims}, we describe the pulsar population characteristics and survey selection effects that are implemented in this study. In Section~\ref{stat_analysis}, we briefly describe the statistical analysis methodology presented in KKL03 and present our results on the individual and total merger rates. In Section~\ref{discuss}, we discuss the implications of our merger rates and compare our total merger rate with that predicted by the LIGO-Virgo group and that estimated through studying the different formation and evolutionary scenarios for DNS systems.
    
\begin{table*}
\centering
\begin{tabular}{rrrrrrrrrrr}
\toprule
& & & & & & & & & \\
{} &          \multicolumn{1}{c}{$l$} &         \multicolumn{1}{c}{$b$} &        \multicolumn{1}{c}{$P$} &        \multicolumn{1}{c}{$\dot{P}$} &       \multicolumn{1}{c}{DM} &       \multicolumn{1}{c}{$P_b$} &       \multicolumn{1}{c}{$a$} &        \multicolumn{1}{c}{$e$} &        \multicolumn{1}{c}{$z$} & \multicolumn{1}{c}{Merger time} \\
PSR        &    (deg)         &     (deg)       &     (ms)      &     ($10^{-18}$ s/s)      &     (pc cm$^{-3}$)     &    (days)      &   (lt-s)       &            &     (kpc)      &       (Gyr)        \\
& & & & & & & & & \\
\midrule
& & & & Non-merging systems & & & & & \\
\midrule
& & & & & & & & & \\
J1518+4904  &   80.8 &  54.3 &  40.9 &  0.027 &   12 &   8.63 &  20.0 &  0.25 &  0.78 &  2400 \\
J0453+1559  &  184.1 & $-$17.1 &  45.8 &  0.19 &   30 &   4.07 &  14.5 &  0.11 & $-$0.15 &  1430 \\
J1811$-$1736  &   12.8 &   0.4 & 104.2 &  0.90 &  476 &  18.78 &  34.8 &  0.83 &  0.03 &  1000 \\
J1411+2551  &   33.4 &  72.1 &  62.4 &  0.096 &   12 &   2.62 &   9.2 &  0.17 &  1.08 &  460 \\
J1829+2456  &   53.3 &  15.6 &  41.0 &  0.052 &   14 &   1.18 &   7.2 &  0.14 &  0.24 &  60 \\
J1753$-$2240  &    6.3 &   1.7 &  95.1 &  0.97 &  159 &  13.64 &  18.1 &  0.30 &  0.09 & - \\
J1930$-$1852  &   20.0 & $-$16.9 & 185.5 &  18.0 &   43 &  45.06 &  86.9 &  0.40 & $-$0.58 &  - \\
& & & & & & & & & & \\
\midrule
& & & & Merging systems & & & & & \\
\midrule
& & & & & & & & & \\
B1534+12    &   19.8 &  48.3 &  37.9 &  2.4 &   12 &   0.42 &   3.7 &  0.27 &  0.79 &  2.70 \\
J1756$-$2251  &    6.5 &   0.9 &  28.5 &  1.0 &  121 &   0.32 &   2.8 & 0.18 &  0.01 &  1.69 \\
J1913+1102  &   45.2 &   0.2 &  27.3 &  0.16 &  339 &   0.21 &   1.7 &  0.09 &  0.02 &  0.50 \\
J1906+0746  &   41.6 &   0.1 &  144.0 &  20000 &  218 &   0.17 &   1.4 &  0.08 &  0.02 &  0.30 \\
B1913+16    &   50.0 &   2.1 &  59.0 &  8.6 &  169 &   0.32 &   2.3 &  0.62 &  0.19 &  0.30 \\
J0737$-$3039A &  245.2 &  $-$4.5 &  22.7 &  1.8 &   49 &   0.10 &   1.4 & 0.09 & $-$0.09 &  0.085 \\
J0737$-$3039B &  245.2 &  $-$4.5 &  2773.5 &  890 &   49 &   0.10 &   1.5 &  0.09 & $-$0.09 &  0.085 \\
J1757$-$1854  &   10.0 &   2.9 &  21.5 &  2.6 &  378 &   0.18 &   2.2 & 0.60 & 0.37 &  0.076 \\
J1946+2052  &   57.7 &  $-$2.0 &  16.9 &     0.90 &   94 &   0.08 &   1.1 &  0.06 &  $-$0.14 &  0.046 \\
\bottomrule
\end{tabular}
\label{psr_table}
\caption{The current sample of DNS systems ranked in decreasing order of time to merger, along with the relevant pulsar and orbital properties. We only consider those systems that will merge within a Hubble time for the merger rate analysis.}
\end{table*}

\section{Pulsar survey simulations}     \label{survey_sims}
    
    To model the pulsar population and survey selection effects, we make use of the freely available PsrPopPy\footnote{\url{https://github.com/devanshkv/PsrPopPy2}} software \citep[][]{psrpoppy, psrpoppy_dag} for generating the population models and writing our own Python code\footnote{\url{https://github.com/NihanPol/2018-DNS-merger-rate}} \citep{code_for_paper, deg_fac_code} to handle all the statistical computation. Here, we describe some of the important selection effects that we model using PsrPopPy.
    
    \subsection{Physical, luminosity and spectral index distribution} \label{physical_dist}
        
        Since we want to calculate the total number of DNS systems like the ones that have been observed, we fix the physical parameters of the pulsars generated in our simulation to represent the DNS systems in which we are interested. These physical parameters include the pulse period, pulse width, and orbital parameters like eccentricity, orbital period, and semi-major axis.
        
        However, even if the physical parameters of the pulsars are the same, their luminosity will not be the same. Thus, to model the luminosity distribution of these pulsars, we use a log-normal distribution with a mean of $\left<\text{log}_{10}L\right> = -1.1$ ($L = 0.07$~mJy~kpc$^2$) and standard deviation, $\sigma_{\text{log}_{10} L} = 0.9$ \citep[][]{FG_kaspi_lum}.
        
        We also vary the spectral index of the simulated pulsar population. We assume the spectral indices have a normal distribution, with mean, $\alpha = -1.4$, and standard deviation, $\beta = 1$ \citep[][]{Bates_si_2013}.
        
    \subsection{Surveys chosen for simulation} \label{surveys}
        
        All of the DNS systems that merge within a Hubble time have either been detected or discovered in the following surveys: the Pulsar Arecibo L-band Feed Array survey \citep[PALFA,][]{PALFA_survey_1}, the High Time-Resolution Universe pulsar survey \citep[HTRU,][]{HTRU_survey_1},  the Parkes High-latitude pulsar survey \citep[][]{PHSURV_1}, the Parkes Multibeam Survey \citep[][]{PMSURV}, and the survey carried out by \citet[][]{1534_disc} in which B1534+12 was discovered. All of these surveys together cover more area on the sky than that covered by the 18 surveys simulated in KKL03 and by \citet[][]{Kim_B_merger}, who included the Parkes Multibeam Pulsar survey in addition to the 18 surveys simulated in KKL03.
        
        We implement these surveys in our simulations with PsrPopPy. We generate a survey file \citep[see Sec.~4.1 in ][]{psrpoppy} for each of these surveys using the published survey parameters. These parameters are then used to estimate the radiometer noise in each survey, which, along with a fiducial signal-to-noise cut-off, will determine whether a pulsar from the simulated population can be detected with a given survey. For example, one important difference in these surveys is their integration time, which ranges from 34~s for Arecibo drift-scan surveys to 2100~s in the Parkes Multibeam survey. Other selection effects can be introduced through differences in the sensitivity of the different surveys, the portion and area of the sky covered and minimum signal-to-noise ratio cut-offs.
        
        PSR J1757--1854 was discovered in the HTRU low-latitude survey using a novel search technique \citep{1757_disc}. As described in \citet[][]{HTRU_search_tech}, the original integration time of 4300~s of the HTRU low-latitude survey was successively segmented by a factor of two into smaller time intervals until a pulsar was detected. This has the effect of reducing Doppler smearing due to extreme orbital motion in tight binary systems (see Sec.~\ref{doppler_smearing} for more on Doppler smearing). The shortest segmented integration time used in their analysis is 537~s (one-eighth segment), which implies that the data are sensitive to binary systems with orbital periods $P_{\rm b} \geq 1.5$~hr \citep{HTRU_search_tech}. All of these segments are searched for pulsars in parallel. We use the integration time of 537~s in our analysis to ensure that the HTRU survey is sensitive to all the DNS systems included in this analysis. We demonstrate the effect of this choice in Sec.~\ref{lum_dist_eff}.
        
        The survey files are available in the GitHub repository associated with this paper.
    
    \subsection{Spatial distribution} \label{spatial_dist}
        
        For the radial distribution of the DNS systems in the Galaxy, like \citet[][]{Kim_B_merger}, we use the model proposed in \citet[][]{Dunc_stats_06}. For the distribution of pulsars in terms of their height, $z$, with respect to the Galactic plane, we use the standard two-sided exponential function \citep[][]{Lyne_stats_1998, Dunc_stats_06},
        \begin{equation}
            \displaystyle f(z) \propto {\rm exp} \left( \frac{-|z|}{z_0} \right)
            \label{z_scale_ht}
        \end{equation}
        where $z_0$ is the vertical scale height. To constrain $z_0$, we simulate DNS populations with a uniform period distribution ranging from 15~ms to 70~ms, consistent with the periods of the recycled pulsars in the DNS systems listed in Table~\ref{psr_table}, and the aforementioned luminosity and spectral index distribution. We generate these populations with vertical scale heights ranging from $z_0 = 0.1$~kpc to $z_0 = 2$~kpc. We run the surveys described in Section~\ref{surveys} on these populations to determine the median vertical scale height of the pulsars that are detected in these surveys. We also calculate the median DM$\times {\rm sin}(|b|)$, which is more robust against errors in converting from dispersion measure to a distance using the NE2001 Galactic electron density model \citep[][]{ne2001}.
        
        We compare these values at different input vertical scale heights with the corresponding median values for the real DNS systems. We show the median DM$\times {\rm sin}(|b|)$ value and the median vertical $z$-height of the pulsars detected in the simulations as a function of the input $z_0$ in Figs.~1~and~2 respectively. In both of these plots, the median values of the real DNS population are plotted as the red dashed line, with the error on the median shown by the shaded cyan region. As can be seen, the analysis using DM$\times {\rm sin}(|b|)$ predicts a vertical scale height of $z_0 = 0.4 \pm 0.1$~kpc, while the analysis using the $z$-height estimated using the NE2001 model \citep{ne2001} returns a vertical scale height of $z_0 = 0.4^{+0.3}_{-0.2}$~kpc. While both these values are consistent with each other, the vertical scale height returned by the DM$\times {\rm sin}(|b|)$ analysis yields a better constraint on the scale height which is more inline with vertical scale heights for ordinary pulsars \citep[$0.33 \pm 0.029$~kpc,][]{z_dist, Dunc_stats_06} and millisecond pulsars \citep[0.5~kpc,][]{Levin_zheight_2013}. We expect the neutron stars that exist in DNS systems, and particularly those DNS systems that merge within a Hubble time, to be born with small natal kicks so as not to disrupt the orbital system. Consequently, we would expect these systems to be closer to the Galactic plane than the general millisecond pulsar population. As a result, we adopt a vertical scale height of $z = 0.33$~kpc as a conservative estimate on the vertical scale height of the DNS population distribution. This difference in the vertical scale height does not result in a significant change in the merger rates.
        
        \begin{figure}
            \centering
            \includegraphics[width = \columnwidth]{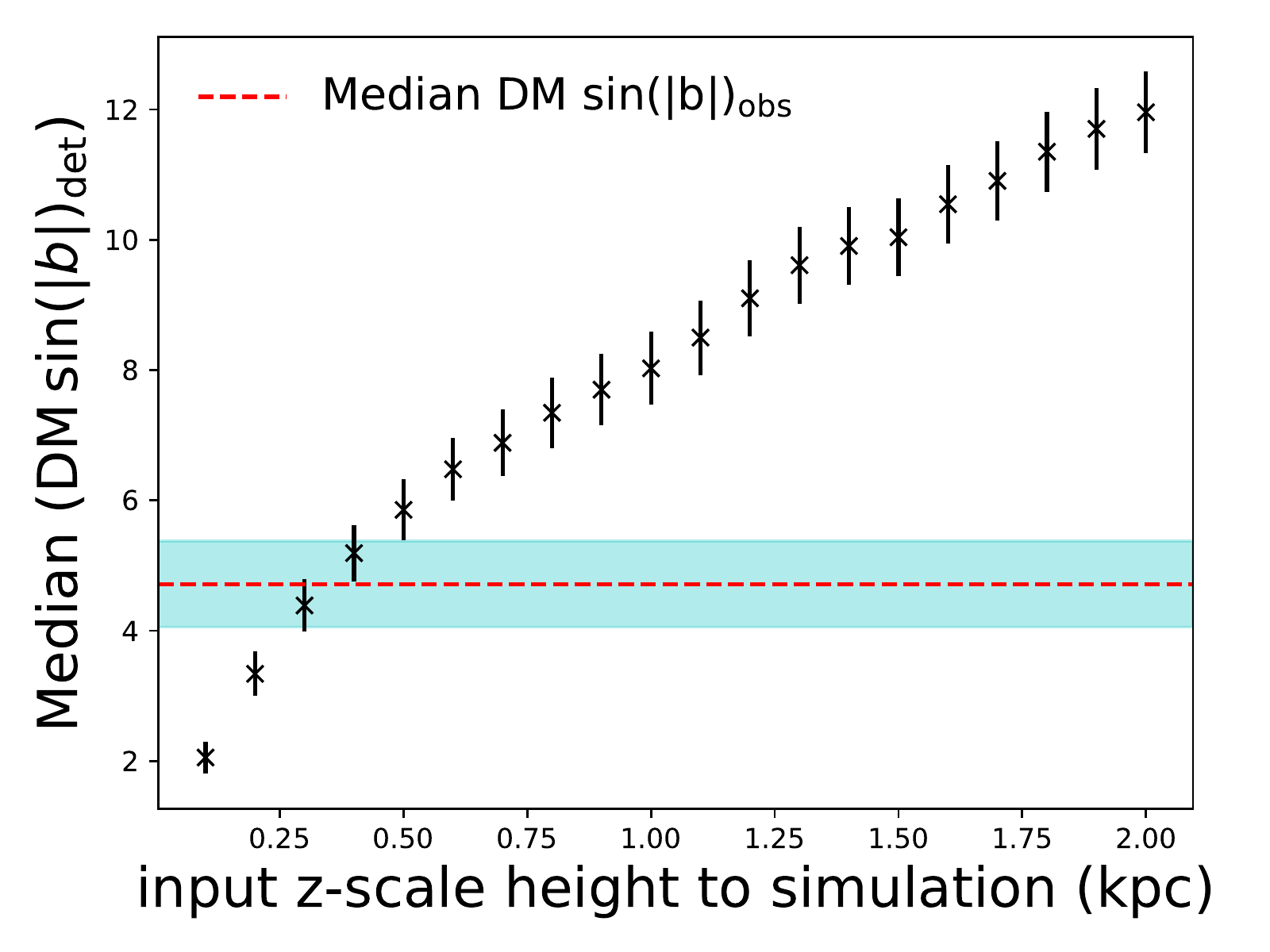}
            \caption{Median observed DM$\times {\rm sin}(|b|)$ plotted versus the input scale height $z_0$ used for the simulated population. The median DM$\times {\rm sin}(|b|)$ value of the real observed population is shown as the horizontal dashed line, with the shaded cyan region depicting 1$\sigma$ error. We can see that populations generated with vertical scale heights ranging from 0.3~kpc to 0.5~kpc agrees with the observed sample.}
            \label{dm_sinb}
        \end{figure}
        
        \begin{figure}
            \centering
            \includegraphics[width = \columnwidth]{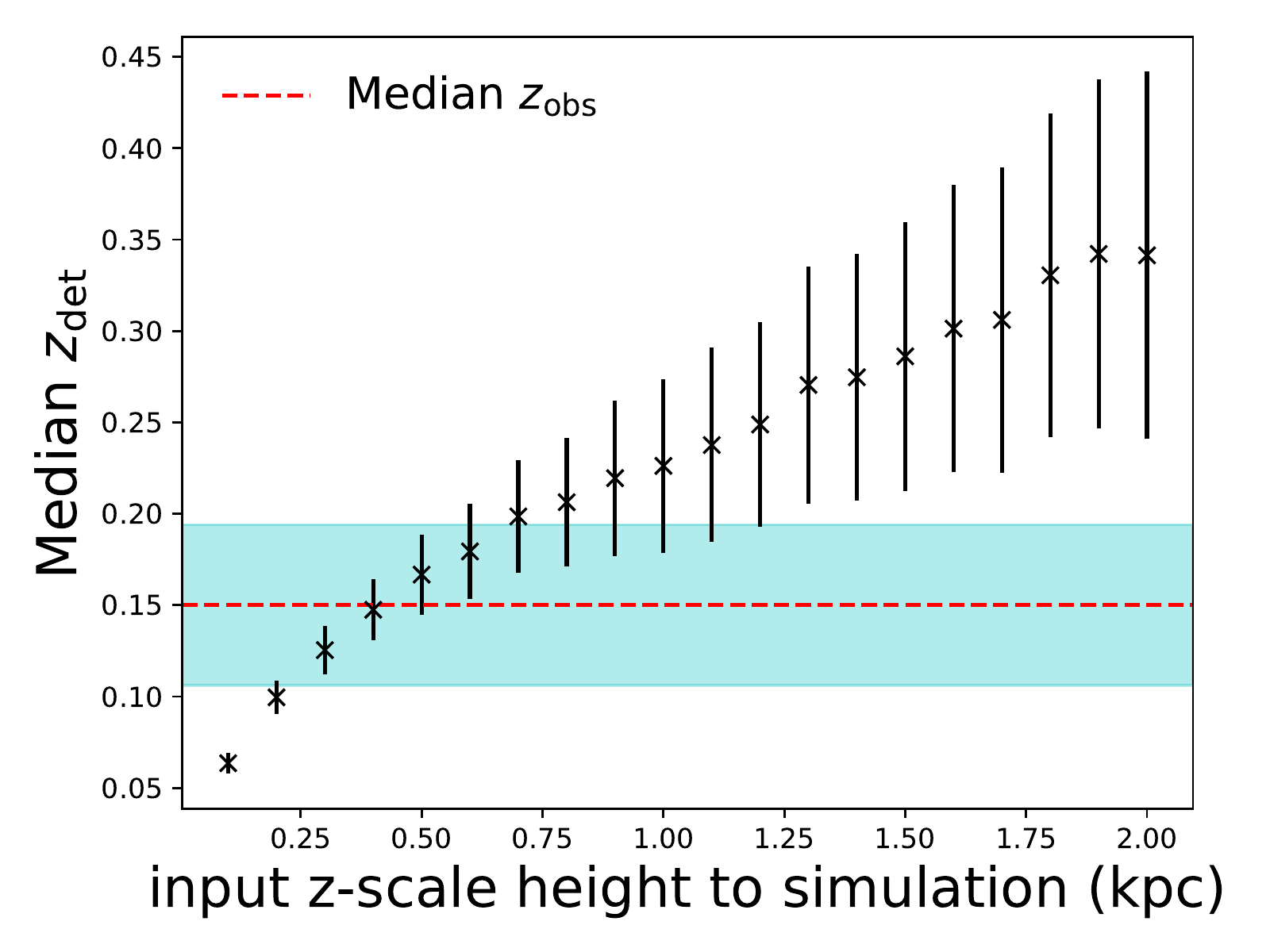}
            \caption{Median vertical scale height calculate using the NE2001 model plotted versus the input scale height $z_0$ used for the simulated population. The median vertical scale height of the real observed population is shown as the horizontal dashed line, with the shaded cyan region depicting 1$\sigma$ error. We can see that populations generated with vertical scale heights ranging from 0.2~kpc to 0.7~kpc agree with the observed sample.}
            \label{z_ht}
        \end{figure}
            
    \subsection{Doppler Smearing} \label{doppler_smearing}
        
        The motion of the pulsar in the orbit of the DNS system introduces a Doppler shifting of the observed pulse period. The extent of the Doppler shift depends on the velocity and acceleration of the pulsar in different parts of its orbit. This Doppler shift results in a reduction in the signal-to-noise ratio for the observation of the pulsar \citep{Bagchi_gamma}.
        
        To account for this effect, we use the algorithm developed by \citet[][]{Bagchi_gamma}, which quantifies the reduction in the signal-to-noise ratio as a degradation factor, $0 < \gamma < 1$, averaged over the entire orbit. This degradation factor depends on the orbital parameters of the DNS system (such as eccentricity and orbital period), the mass of the two neutron stars, the integration time for the observation, and the search technique used in the survey \citep[for example, HTRU and PALFA surveys use acceleration-searches;][]{Bagchi_gamma}. A degradation factor $\gamma \sim 1$ implies very little Doppler smearing, while a degradation factor $\gamma \sim 0$ implies heavy Doppler smearing in the pulsar's radio emission.
        
        The implementation of the algorithm as a Fortran program was kindly provided to us by the authors of \citet[][]{Bagchi_gamma}, which we make available\footnote{\url{https://github.com/NihanPol/SNR\_degradation\_factor\_for\_BNS\_systems}} with their permission. Since PsrPopPy does not include functionality to handle this degradation factor, we had to manually introduce the degradation factor into the source code of PsrPopPy. The modified PsrPopPy source files are also available on the GitHub repository.
            
    \subsection{Beaming correction factor} \label{beaming_fraction}
        
        The beaming correction fraction, $f_b$, is defined as the inverse of the pulsar's beaming fraction, i.e. the solid angle swept out by the pulsar's radio beam divided by $4 \pi$. PSRs B1913+16, B1534+12, and J0737--3039A/B have detailed polarimetric observation data, from which precise measurement of their beaming fractions, and thus, their beaming corrections factors has been possible. These beaming corrections are collected in Table~2 of \citet[][]{Kim_B_merger}.
        
        However, the other merging DNS systems are relatively new discoveries and do not have measured values for their beaming fractions. Thus, we assume that the beaming correction factor for these new pulsars is the average of the measured beaming correction factor for the three aforementioned pulsars, i.e. 4.6. We list these beaming fractions in Table~\ref{result_table}, and defer discussion of their effect on the merger rate for Section~\ref{discuss}.
        
    \subsection{Effective lifetime} \label{eff_life}
        
        The effective lifetime of a DNS binary, $\tau_{\rm life}$, is defined as the time interval during which the DNS system is detectable. Thus, it is the sum of the time since the formation of the DNS system and the remaining lifetime of the DNS system,
        \begin{multline}
            \displaystyle \tau_{\rm life} = \tau_{\rm age} + \tau_{\rm obs} \\
             = {\rm min} \left( \tau_{\rm c}, \tau_{\rm c}\left[1 - \frac{P_{\rm birth}}{P_{\rm s}} \right]^{n - 1} \right) + {\rm min}(\tau_{\rm merger}, \tau_{\rm d}).
            \label{lifetime}
        \end{multline}
        Here $\tau_{\rm c} = P_{\rm s} / (n - 1) \dot{P_{\rm s}}$ is the characteristic age of the pulsar, $n$ is the braking index, assumed to be 3, $P_{\rm birth}$ is the period of the millisecond pulsar at birth, i.e.~when it begins to move away from the fiducial spin-up line on the $P-\dot{P}$ diagram, $P_{\rm s}$ is the current spin period of the pulsar, $\tau_{\rm merger}$ is the time for the DNS system to merge, and $\tau_{\rm d}$ is the time in which the pulsar crosses the ``death line'' beyond which pulsars should not radiate significantly \citep[][]{death_line_1}.
        
        Unlike normal pulsars, the characteristic age, $\tau_{\rm c}$, for millisecond and recycled pulsars may not be a very good indicator of the true age of the pulsar. This is due to the fact that the period of the pulsar at birth is much smaller than the current period of the pulsar, which is not true for recycled millisecond pulsars found in DNS systems. A better estimate for the age of a recycled millisecond pulsar can be calculated by measuring the distance of the pulsar from a fiducial spin-up line on the $P-\dot{P}$ diagram \citep[][]{true_age}, represented by the second part of the first term in Eq.~\ref{lifetime}.
        
        Finally, the time for which a given DNS system is detectable after birth depends on whether we are observing the non-recycled companion pulsar (J0737--3039B, J1906+0746) or the recycled pulsar in the DNS system (e.g.~B1913+16, J1757--1854, J1946+2052, etc.). In the latter case, the combination of a small spin-down rate and millisecond period ensures that the DNS system remains detectable until the epoch of the merger. However, for the former case, both the period and spin-down rate are at least an order of magnitude larger than their recycled counterparts. As such, the time for which these systems are detectable depends on whether they cross the pulsar ``death line" before their epoch of merger \citep[][]{death_line_1}. The radio lifetime of any pulsar is defined as the time it takes the pulsar to cross this fiducial ``death line'' on the $P-\dot{P}$ diagram \citep{death_line_1}.
        
        We estimate the radio lifetime for J1906+0746 using two different techniques. One estimate is described by \citet[][]{death_line_1} and assumes a simple dipolar rotator to find the time to cross the deathline. Using Eq.~6 in their paper we calculate a radio lifetime of $\tau_{\rm d} \sim 3$~Myr for J1906+0746. However, as discussed in \citet[][]{death_line_1}, the death line for a pure dipolar rotator might not be an accurate turn-off point for pulsars, with many observed pulsars lying past this line on the $P-\dot{P}$ diagram. A better estimate of the radio lifetime might be given by Eq.~9 in \citet[][]{death_line_1}, which assumes a twisted field configuration for pulsars. Using this, we find $\tau_{\rm d} \sim 30$~Myr. Another estimate for the radio lifetime can be made from spin-down energy loss considerations. Adopting the formalism given in \citet[][]{death_line_2}, we find, for a simple dipolar spin-down model, that a pulsar with a current spin-down energy loss rate $\dot{E}_{\rm now}$ and characteristic age $\tau$ will reach a cut-off $\dot{E}$ value of $10^{30}$~ergs/s below which radio emission through pair production is suppressed on a timescale
        \begin{equation}
            \displaystyle \tau_{\rm d} = \tau_{\rm c} \left( \sqrt{\frac{\dot{E}_{\rm now}}{\dot{E} = 10^{30}}} - 1 \right).
            \label{death_line_1}
        \end{equation}
        Using this formalism, we calculate a radio lifetime of $\tau_{\rm d} \sim 60$~Myr. This method of estimation has been used in previous estimates \citep{comp_merger_rate_analysis, dunc_merger_1, Kim_B_merger} of the merger rates, and represents a conservative estimate on the radio lifetime of J1906+0746. We adopt it here as the fiducial radio lifetime of J1906+0746 for consistency, and defer the discussion of the implications of variation in the calculated radio lifetime to Sec.~\ref{discuss}.
        
        A similar analysis could be done for pulsar B in the J0737--3039 system. However, unlike \citet[][]{Kim_B_merger}, we do not include B in our merger rate calculations. The uncertainties in the radio lifetime are very large, as for PSR J1906+0746, and therefore pulsar A provides a much more reliable estimate of the numbers of such systems. In addition, unlike J1906+0746, pulsar B also shows large variations in its equivalent pulse width \citep{Kim_B_merger}, and thus, its duty cycle, due to pulse profile evolution through geodetic precession \citep{B_pulse_ev_perera}. This also leads to an uncertainty in its beaming correction factor \citep[see Fig.~4 in ][]{Kim_B_merger}. There are additional uncertainties introduced by pulsar B exhibiting strong flux density variations over a single orbit around A. All these factors introduce a large uncertainty in the merger rate contribution from B, and do not provide better constraints on the merger rate compared to when only pulsar A is included \citep{Kim_B_merger}. Finally, the Double Pulsar system was discovered through pulsar A and will remain detectable through pulsar A long after B crosses the death line. Due to these reasons, we do not include pulsar B in our analysis.
    
\section{Statistical Analysis and Results}      \label{stat_analysis}
    
    Our analysis is based on the procedure laid out in \citet[][]{dunc_merger_1} (hereafter KKL03). For completeness, we briefly outline the process below.
    
    We generate populations of different sizes, $N_{\rm tot}$, for each of the known, merging DNS systems which are beaming towards us in physical and radio luminosity space using the observed pulse periods and pulse widths. The choice of the physical and luminosity distribution is discussed in Sec.~\ref{physical_dist}. On each population, we run the surveys described in Sec.~\ref{surveys} to determine the total number of pulsars that will be detected, $N_{\rm obs}$, in those surveys. The population size, $N_{\rm tot}$, that returns a detection of one pulsar, i.e. $N_{\rm obs} = 1$, will represent the true size of the population of that DNS system.
    
    For a given $N_{\rm tot}$ pulsars of some type in the Galaxy, and the corresponding $N_{\rm obs}$ pulsars that are detected, we expect the number of observed pulsars to follow a Poisson distribution:
    \begin{equation}
        \displaystyle P(N_{\rm obs}; \lambda) = \frac{\lambda^{N_{\rm obs}} e^{-\lambda}}{N_{\rm obs}!}
        \label{poisson}
    \end{equation}
    where, by definition, $\lambda \equiv \left< N_{\rm obs} \right>$. Following arguments presented in KKL03, we know that the linear relation
    \begin{equation}
        \displaystyle \lambda = \alpha N_{\rm tot}
        \label{alpha_eq}
    \end{equation}
    holds. Here $\alpha$ is a constant that depends on the properties of each of the DNS system populations and the pulsar surveys under consideration.
    
    The likelihood function, $P(D|HX)$, where $D = 1$ is the real observed sample, $H$ is our model hypothesis, i.e. $\lambda$ which is proportional to $N_{\rm tot}$, and $X$ is the population model, is defined as:
    \begin{equation}
        \displaystyle P(D|HX) = P(1|\lambda(N_{\rm tot}), X) = \lambda(N_{\rm tot}) e^{-\lambda(N_{\rm tot})}
        \label{likelihood_fn}
    \end{equation}
    Using Bayes' theorem and following the derivation given in KKL03, the posterior probability distribution, $P(\lambda| DX)$, is equal to the likelihood function. Thus,
    \begin{equation}
        \displaystyle P(\lambda| DX) \equiv P(\lambda) = P(1|\lambda(N_{\rm tot}), X) = \lambda(N_{\rm tot}) e^{-\lambda(N_{\rm tot})}.
        \label{posterior_fn}
    \end{equation}
    Using the above posterior distribution function, we can calculate the probability distribution for $N_{\rm tot}$,
    \begin{equation}
        \displaystyle P(N_{\rm tot}) = P(\lambda) \left| \frac{d\lambda}{dN_{\rm tot}} \right| = \alpha^2 N_{\rm tot} e^{-\alpha N_{\rm tot}}
        \label{ntot_prob}
    \end{equation}
    For a given total number of pulsars in the Galaxy, we can calculate the corresponding Galactic merger rate, $\cal{R}$, using the beaming fraction, $f_{\rm b}$, of that pulsar and its lifetime, $\tau_{\rm life}$, as follows:
    \begin{equation}
        \displaystyle \mathcal{R} = \frac{N_{\rm tot}}{ \tau_{\rm life}} f_{\rm b}.
        \label{rate_eq}
    \end{equation}
    Finally, we calculate the Galactic merger rate probability distribution
    \begin{equation}
        \displaystyle P(\mathcal{R}) = P(N_{\rm tot}) \frac{dN_{\rm tot}}{d\mathcal{R}} = \left( \frac{\alpha\, \tau_{\rm life}}{f_b} \right)^2                                 \mathcal{R} \, e^{-(\alpha \tau_{\rm life} / f_b)\mathcal{R}}.
        \label{rate_pdf}
    \end{equation}
    
    \begin{figure*}[htb]
        \centering
        \includegraphics[width = \textwidth]{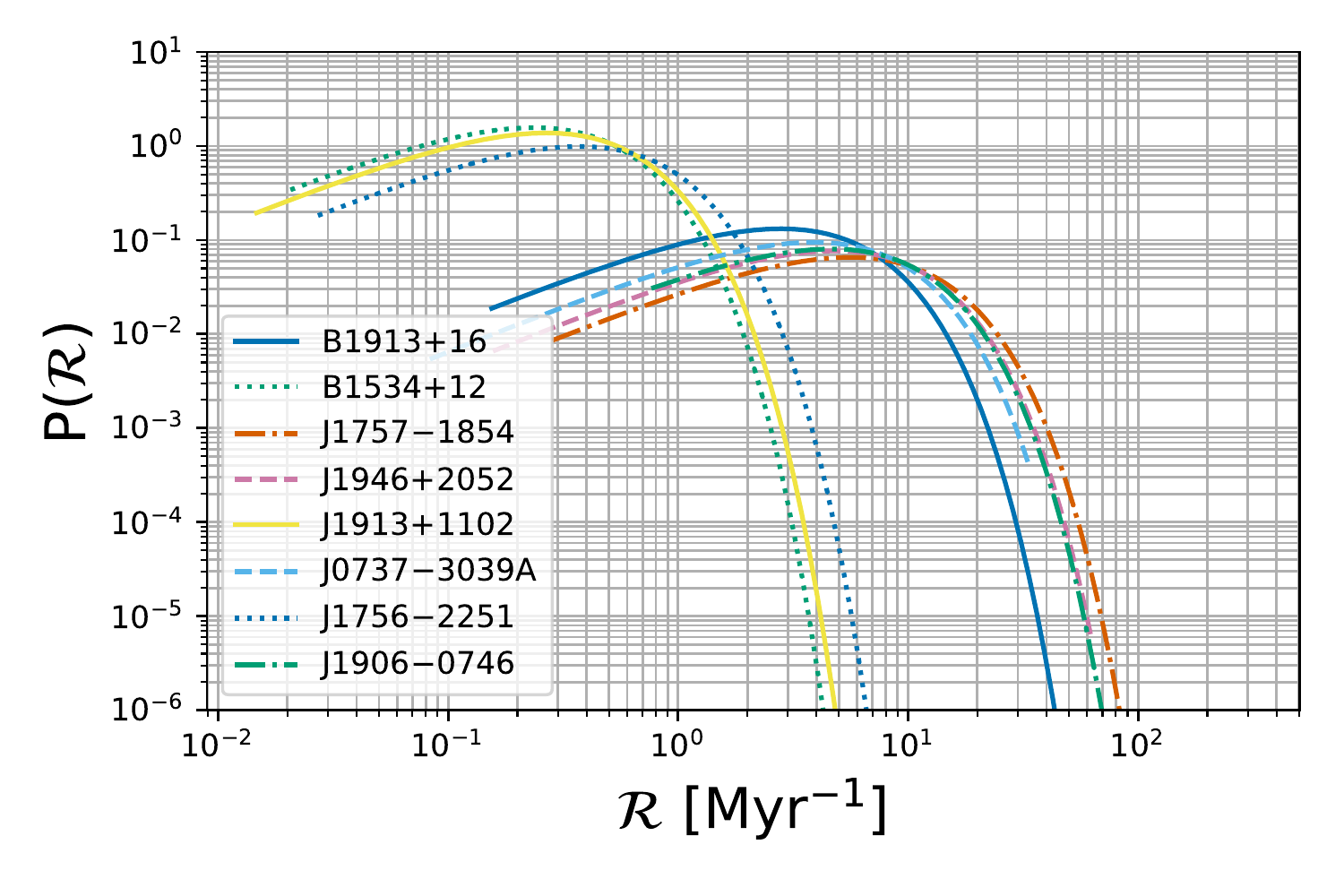}
        \caption{Probability distribution function of the Galactic merger rate of each individual DNS system. We can see that J1757--1854, J1946+2052, J0737--3039 and B1913+16 have the largest individual merger rates, followed by J1906+0746 and then B1534+12, J1756--2251 and J1913+1102.}
        \label{individual_rates}
    \end{figure*}
    
    Following the above procedure for all the merging DNS systems, we obtain the individual Galactic merger rates for each system, which are shown in Fig.~\ref{individual_rates}.
    
    \subsection{Calculating the total Galactic merger rate}
                
        After calculating individual merger rates from each DNS system, we need to combine these merger rate probability distributions to find the combined Galactic probability distribution. We can do this by treating the merger rate for the individual DNS systems as independent continuous random variables. In that case, the total merger rate for the Galaxy will be the arithemtic sum of the individual merger rates
        \begin{equation}
            \displaystyle \mathcal{R}_{\rm MW} = \sum_{i = 1}^{8} \mathcal{R}_{i}
            \label{gal_rate}
        \end{equation}
        with the total Galactic merger rate probability distribution given by a convolution of the individual merger rate probability distributions,
        \begin{equation}
            \displaystyle P(\mathcal{R}_{\rm MW}) = \prod_{i = 1}^{8}  P(\mathcal{R}_{i})
            \label{total_gal_rate}
        \end{equation}
        where $\prod$ denotes convolution. As the number of known DNS systems increases over time, the method of convolution of individual merger rate PDFs is more efficient than computing an explicit analytic expression as in KKL03 and \citet[][]{Kim_B_merger}.
        
        \begin{figure}[htb]
            \centering
            \includegraphics[width = \columnwidth]{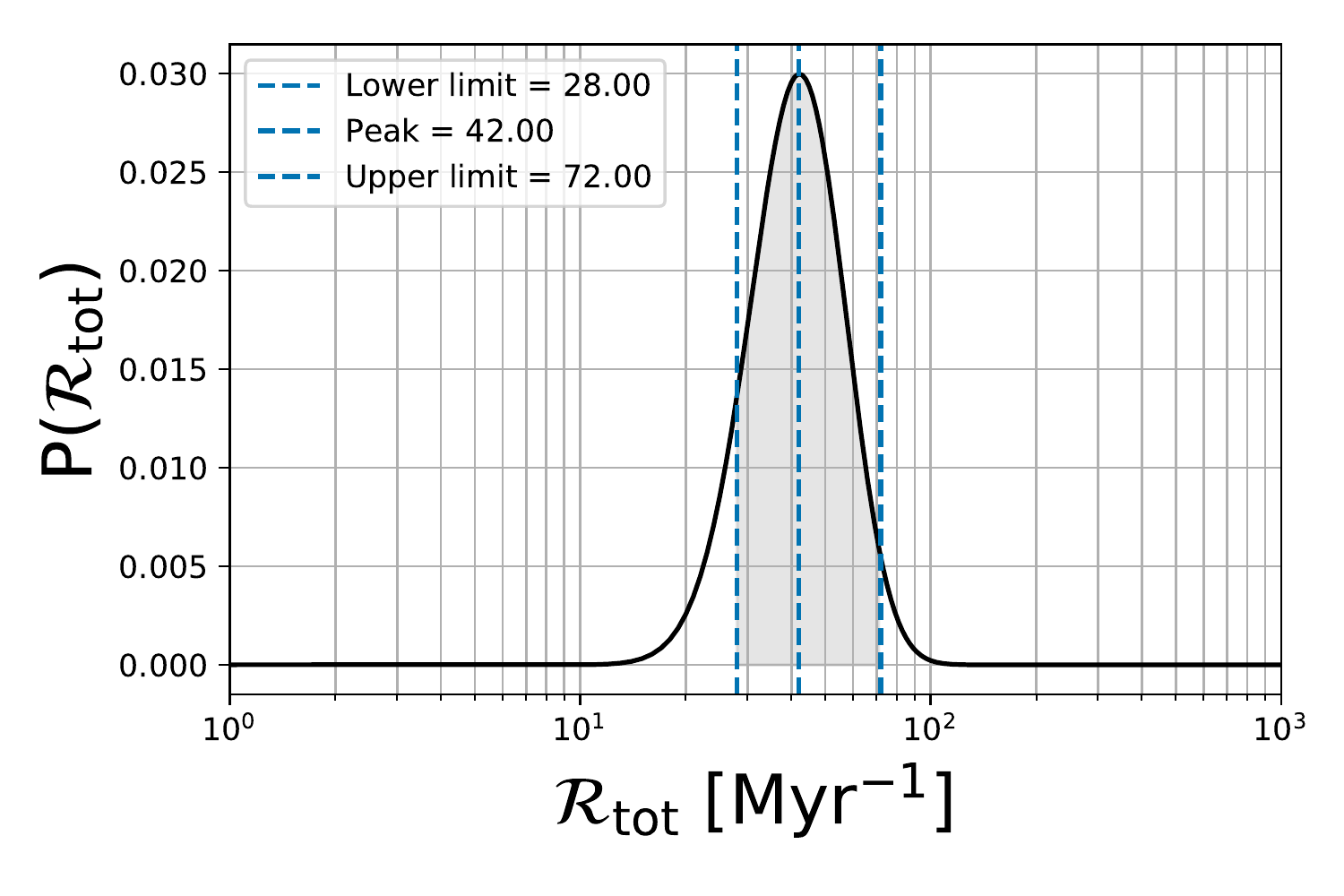}
            \caption{The total Milky Way DNS merger rate probability distribution function. This distribution is obtained by convolution of the indivual merger rate probability distributions as described in Section~\ref{stat_analysis}. The shaded region denotes the 90\% confidence interval, which was calculated starting from the peak of the distribution and collecting 45\% probability on both sides of the peak independently, while the vertical dashed lines represent the limits on the Milky Way DNS merger rate at the 90\% confidence interval.}
            \label{total_mw_rate}
        \end{figure}
        
        Combining all the individual Galactic merger rates, we obtain a total Galactic merger rate of $\mathcal{R}_{\rm MW} = 42^{+30}_{-14}$~Myr$^{-1}$, which is shown in Fig.~\ref{total_mw_rate}.
        
    \subsection{The merger detection rate for advanced LIGO}
        
        The Galactic merger rate calculated above can be extrapolated to calculate the number of DNS merger events that LIGO will be able to detect. Assuming that the DNS formation rate is proportional to the formation rate of massive stars, which is in turn proportional to the $B$-band luminosity of a given galaxy \citep[][]{Phinney_blue_lum_scaling, comp_merger_rate_analysis}, the DNS merger rate within a sphere of radius $D$ is given by \citep[][]{extrapolate_to_get_LIGO_rate}
        \begin{equation}
            \displaystyle \mathcal{R}_{\rm LIGO} = \mathcal{R}_{\rm MW} \left( \frac{L_{\rm total} (D)}{L_{\rm MW}} \right)
            \label{b_band_luminosity}
        \end{equation}
        where $L_{\rm total} (D)$ is the total blue luminosity within a distance $D$, and $L_{\rm MW} = 1.7 \times 10^{10} L_{B, \odot}$, where $L_{B, \odot} = 2.16 \times 10^{33}$~ergs/s, is the $B$-band luminosity of the Milky Way \citep[][]{extrapolate_to_get_LIGO_rate}.
        
        Using a reference LIGO range distance of $D_{\rm r} = 100$~Mpc \citep[][]{LIGO_horizon_dist}, and following the arguments laid out in \citet[][]{extrapolate_to_get_LIGO_rate}, we can calculate the rate of DNS merger events visible to LIGO \citep[equation 19 in ][]{extrapolate_to_get_LIGO_rate}
        \begin{multline}
            \displaystyle \mathcal{R}_{\rm LIGO} = \frac{N}{T} \\
            = 7.4 \times 10^{-3} \left( \frac{\mathcal{R}}{(10^{10} L_{B, \odot})^{-1}\,{\rm Myr}^{-1}} \right) \left( \frac{D_{\rm r}}{100\,{\rm Mpc}} \right)^3 \rm yr^{-1}
            \label{extrapol}
        \end{multline}
        where $N$ is the number of mergers in $T$ years, and $\mathcal{R} = \mathcal{R_{\rm MW}} / L_{\rm MW}$ is the Milky Way merger rate weighted by the Milky Way B-band luminosity.
        
        Using the above equation, we calculate the DNS merger detection rate for LIGO,
        \begin{equation}
            \displaystyle \mathcal{R}_{\rm LIGO} \equiv \mathcal{R}_{\rm PML18} = 0.18^{+0.13}_{-0.06} \times \left( \frac{D_{\rm r}}{100 \ \rm Mpc} \right)^3 \rm yr^{-1},
            \label{ligo_rate}
        \end{equation}
        where we use $\mathcal{R}_{\rm PML18}$ to distinguish our merger detection rate estimate from the others that will be referred to later in the paper.
    
\section{Discussion}        \label{discuss}
    
    \begin{table*}[]
        \centering
        \begin{tabular}{ccccccc}
            \toprule
            PSR & $f_b$ & $\delta$ & $\tau_{\rm age}$ & $N_{\rm obs}$ & $N_{\rm pop}$ & $\mathcal{R}$ \\
             & & & (Myr) & & & (Myr$^{-1}$) \\
            \midrule
            & & & & & & \\
            B1534+12 & 6.0 & 0.04 & 208 & $114^{+519}_{-80}$ & $688^{+2546}_{-483}$ & $0.2^{+1.1}_{-0.1}$ \\
            & & & & & & \\
            J1756$-$2251 & 4.6 & 0.03 & 396 & $138^{+627}_{-96}$ & $633^{+2878}_{-440}$ & $0.4^{+1.7}_{-0.3}$ \\
            & & & & & & \\
            J1913+1102 & 4.6 & 0.06 & 2625 & $182^{+830}_{-132}$ & $834^{+3813}_{-605}$ & $0.3^{+1.2}_{-0.2}$ \\
            & & & & & & \\
            J1906+0746 & 4.6 & 0.01 & 0.11 & $62^{+276}_{-36}$ & $284^{+1265}_{-165}$ & $4.7^{+21.1}_{-2.7}$ \\
            & & & & & & \\
            B1913+16 & 5.7 & 0.169 & 77 & $182^{+822}_{-132}$ & $1040^{+4706}_{-754}$ & $2.8^{12.3}_{-2.0}$ \\
            & & & & & & \\
            J0737$-$3039A & 2.0 & 0.27 & 159 & $465^{+2097}_{-347}$ & $931^{+4193}_{-695}$ & $3.9^{+17.4}_{-2.9}$ \\
            & & & & & & \\
            J1757$-$1854 & 4.6 & 0.06 & 87 & $198^{+906}_{-144}$ & $908^{+4161}_{-660}$ & $5.6^{+25.7}_{-4.1}$ \\
            & & & & & & \\
            J1946+2052 & 4.6 & 0.06 & 247 & $306^{+1397}_{-224}$ & $1402^{+6417}_{-1026}$ & $4.8^{+21.9}_{-5.5}$ \\
            & & & & & & \\
            \bottomrule
        \end{tabular}
        \caption{Parameters and results from the DNS merger rate analysis described in Section~\ref{stat_analysis}. Here $f_b$ is the beaming correction factor, $\delta$ is the pulse duty cycle, $\tau_{\rm age}$ is the effective age described in Section~\ref{lifetime}, $N_{\rm obs}$ is the number of each DNS system that are beaming towards the Earth, $N_{\rm pop}$ is the total number of each DNS system in the Milky Way, and $\mathcal{R}$ is the merger rate of each individual DNS system population, the probability distribution function for which is shown in Fig.~\ref{individual_rates}. The errors on the quantities represent 95\% confidence interval.}
        \label{result_table}
    \end{table*}
    
    In this paper, we consider eight DNS systems that merge within a Hubble time, and using the procedure described in KKL03 estimate the Galactic DNS merger rate to be $\mathcal{R}_{\rm MW} = 42^{+30}_{-14}$~Myr$^{-1}$ at 90\% confidence. This is a modest increase from the most recent rate calculated by \citet[][$\mathcal{R}_{\rm MW} = 21^{+28}_{-14}$~Myr$^{-1}$ at the 95\% confidence level]{Kim_B_merger}, despite the addition of five new DNS systems in our analysis. This is due to the addition of two large scale surveys (the PALFA and HTRU surveys) to our analysis, as a result of which we are sampling a significantly larger area on the sky than \citet[][]{Kim_B_merger}. This larger fraction of the sky surveyed, coupled with only a few new DNS discoveries, contributes to the overall reduction in the population of the individual DNS systems. For example, \citet[][]{Kim_B_merger} predict that there should be $\sim$907 J0737--3039A-like systems in the galaxy, while our analysis predicts a lower value of $\sim$465 such systems. This reduced population of individual DNS systems leads to a reduction in their respective contribution to the merger rate. 
    
    Irrespective of the reduction in the individual DNS system population, the five new DNS systems added in this analysis cause an overall increase in the Galactic merger rate. As shown in Fig.~\ref{individual_rates}, J1757--1854, J1946+2052 and J1906+0746 have the highest contributions to the merger rate along with J0737--3039 and B1913+16, while the other two DNS systems of J1913+1102 and J1756--2251 round out the Galactic merger rate with relatively smaller contributions. We do not consider pulsar B from the J0737--3039 system in our analysis. The inclusion of pulsar A is sufficient to model the contribution of the Double Pulsar to the merger rates \citep[][]{Kim_B_merger} and the inclusion of B does not lead to a better constraint on the merger rate.
    
    \subsection{Comparison with the LIGO DNS merger detection rate}
        
        The recent detection of a DNS merger by LIGO \citep[][]{THE_DNS_merger} enabled a calculation of the rate of DNS mergers visible to LIGO \citep[][]{THE_DNS_merger}. The rate that was calculated in \citet[][]{THE_DNS_merger}, converted to the units used in our calculations, is
        \begin{equation}
            \displaystyle \mathcal{R}_{\rm LIGO} \equiv \mathcal{R}_{\rm A17} = 1.54^{+3.20}_{-1.22} \times \left( \frac{D_{\rm r}}{100 \ \rm Mpc} \right)^3 \rm yr^{-1}
            \label{LIGO_rate}
        \end{equation}
        where $\mathcal{R}_{\rm A17}$ is the merger detection rate and the errors quoted are 90\% confidence intervals.
        
\begin{figure*}[htb]
            \centering
            \includegraphics[width = \textwidth]{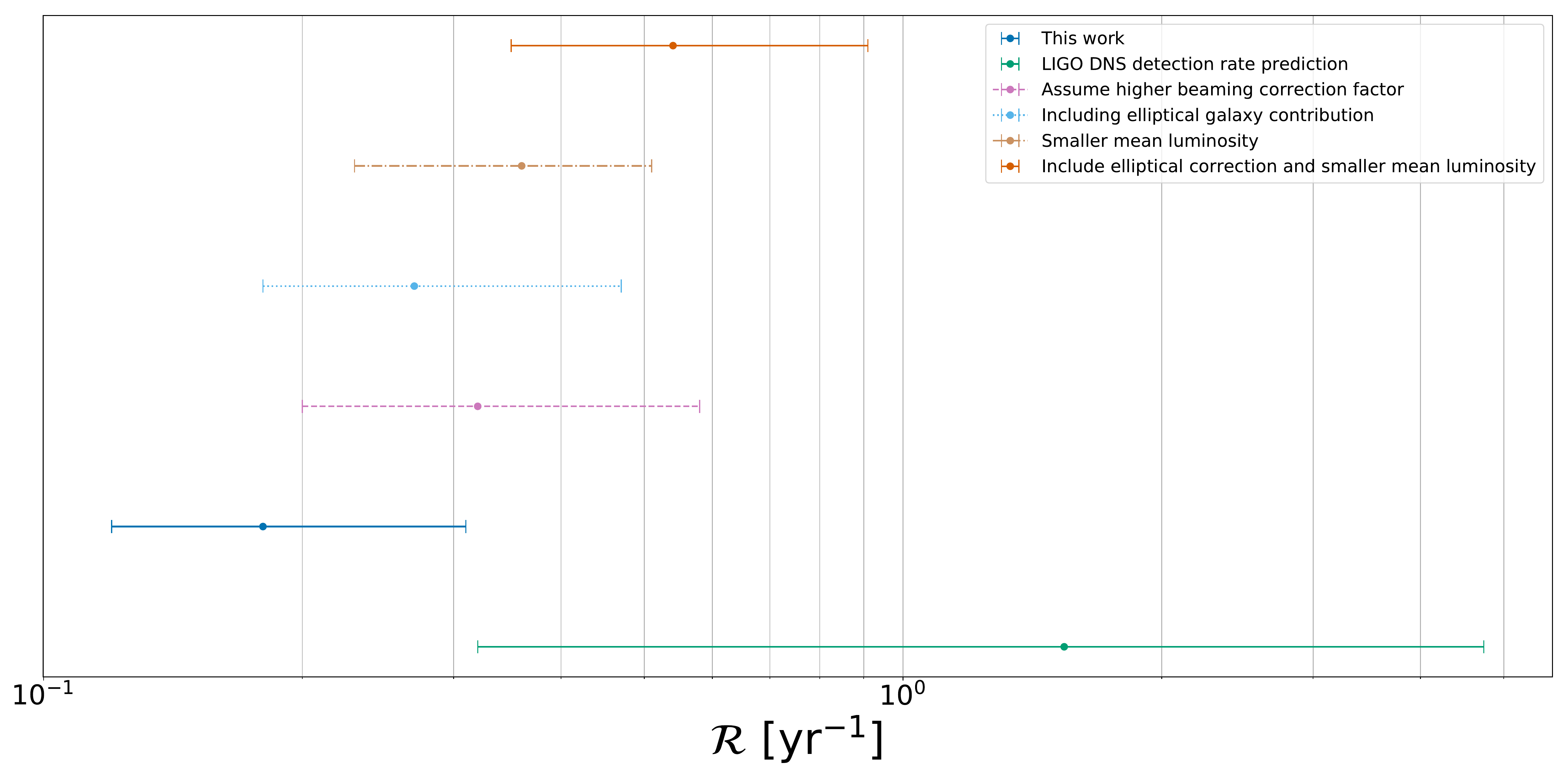}
            \caption{We compare the variation in the merger detection rate calculated in this work with change in the different underlying assumptions used in the derivation of the rate. We show the effects of variation in the luminosity distribution of the DNS population, assuming a large beaming correction factor, and including the contribution of elliptical galaxies in LIGO's observable volume (see text for details). We also plot the modified merger detection rate that includes both the correction for elliptical galaxies and a fainter DNS population. We do not include in this the overestimated beaming correction factor effect as we do not think the beaming correction factors will differ significantly from those assumed in this work.}
            \label{var_sim}
        \end{figure*}

        We plot both the rate estimates in Figure~\ref{var_sim}. This rate estimated by LIGO is in agreement with the DNS merger detection rate that we calculate using the Milky Way DNS binary population, $\mathcal{R}_{\rm PML18}$, at the upper end of the 90\% confidence level range.
        
    \subsection{Caveats on our merger and detection rates}
        
        \subsubsection{Luminosity distribution} \label{lum_dist_eff}
            
            In generating the populations of each type of DNS system in the Galaxy, we assumed a log-normal distribution with a mean of $\left<\text{log}_{10}L\right> = -1.1$ and standard deviation, $\sigma_{\text{log}_{10} L} = 0.9$ \citep[][]{FG_kaspi_lum}. This distribution was found to adequately represent ordinary pulsars by \citet[][]{FG_kaspi_lum}. However, the DNS system population might not be well represented by this distribution. The dearth of known DNS systems prevents an accurate measurement of the mean and standard deviation of the log-normal distribution for the DNS population.
            
            The sample of DNS systems in the Galaxy might be well represented by the sample of recycled pulsars in the Galaxy. \citet[][]{Bagchi_gc_lum_func} analyzed the luminosity distribution of the recycled pulsars found in globular clusters, and concluded that both powerlaw and log-normal distributions accurately model the observed luminosity distribution, though there was a wide spread in the best-fit parameters for both distributions. They found that the luminosity distribution derived by \citet[][]{FG_kaspi_lum} is consistent with the observed luminosity distribution of recycled pulsars.
            
            We also assumed an integration time for the HTRU low-latitude survey (537~s) that is one-eighth of the integration time of the survey (4300~s) \citep[see Sec.~\ref{surveys} and][]{HTRU_search_tech}. Based on the radiometer equation, this implies a reduction in sensitivity by a factor of $\sim 2.8$ \citep{psr_handbook} in searching for a given pulsar.
            
            To test the effect of the above on $\mathcal{R}_{\rm PML18}$, we used the results from \citet[][]{Bagchi_gc_lum_func} to pick a set of parameters for the log-normal distribution that represents a fainter population of DNS systems in the Galaxy. We pick a mean of $\left<\text{log}_{10}L\right> = -1.5$ (consistent with the lower flux sensitivity of the HTRU low-latitude survey) and standard deviation, $\sigma_{\text{log}_{10} L} = 0.94$ \citep[][]{Bagchi_gc_lum_func}. This increases our merger detection rate to $0.36^{+0.15}_{-0.13}$~yr$^{-1}$, which is a factor of 1.5 larger than our calculated merger detection rate. This demonstrates that if the DNS population is fainter than the ordinary pulsar population, we would see a marked increase in the merger detection rate.
            
        \subsubsection{Beaming correction factors}
            
            In our analysis, we use the average of the beaming correction factors measured for B1913+16, B1534+12, and J0737--3039A (see Table~\ref{result_table}) as the beaming correction factors for the newly added DNS systems. However, the Milky Way merger rate that we calculate is sensitive to changes in the beaming correction factors for the newly added DNS systems. To demonstrate this, we changed the beaming correction factors of all the new DNS systems to 10, i.e. slightly more than twice the values that we use. The resulting merger detection rate then increases to $0.32^{+0.26}_{-0.12}$~yr$^{-1}$, which is a 77.77\% increase from the original merger detection rate $\mathcal{R}_{\rm PML18}$. 
            
            Even though this is a significant increase in the merger detection rate, it is highly unlikely to see beaming correction factors as large as 10. The study by \citet[][]{beaming_fraction_review_1} demonstrates that pulsars with periods between 10~ms $< P <$ 100~ms are likely to have beaming correction factors $\sim 6$, with predictions not exceeding 8 in the most extreme cases \citep[see Figs.~3~and~4 in][]{beaming_fraction_review_1}. As a result, we do not expect a huge change in the merger detection rate due to variations in the beaming correction factors for the new DNS systems added in this analysis.
            
        \subsubsection{The effective lifetime of J1906+0746}
            
            PSR~J1906+0746 is an interesting DNS system which highlights the significance of the effective lifetime in the Galactic merger rate and the merger detection rate calculations. The properties of J1906+0746 suggest that it is similar to pulsar B in the Double Pulsar system. However, all searches for a companion pulsar in the J1906+0746 system have been negative \citep{1906_dns_evidence}. Just like J0737--3039B, the combination of a long period and high period derivative implies that the radio lifetime of J1906+0746 might be shorter than the coalescence timescale of the system through emission of gravitational waves. 
            
            As shown in Sec.~\ref{eff_life}, there is more than an order of magnitude variation in the estimated radio lifetime of J1906+0746. Including the gravitational wave coalescence timescale, the range of possible radio lifetimes, and hence the effective lifetimes (the characteristic age of J1906+0746 is a tender 110~kyr), for J1906+0746 ranges from $3~{\rm Myr} < \tau_{\rm eff} < 300~{\rm Myr}$. This has a significant impact on the contribution of J1906+0746 towards the merger detection rate through Eq.~\ref{rate_eq}, and thus the complete merger detection rate. For example, if $\tau_{\rm d} = 3$~Myr is an accurate estimate of the effective lifetime of J1906+0746, our merger detection rate would increase to $0.57^{+1.52}_{-0.24}$. In this scenario, J1906+0746 would contribute as much as $\sim 95$~Myr$^{-1}$ to the Galactic merger rate, compared to its contribution of $\sim 5$~Myr$^{-1}$ in the fiducial scenario. However, as pointed out earlier, it is unlikely that the effective lifetime of J1906+0746 will be as short as $\tau_{\rm d} = 3$~Myr. At the other extreme, an effective lifetime of $\tau_{\rm d} = 300$~Myr would reduce our merger detection rate to $0.15^{+0.12}_{-0.05}$. This effective lifetime is almost certainly longer than the true effective lifetime of J1906+0746 by about an order of magnitude as shown by the different calculations in Sec.~\ref{eff_life}.
            
            Thus, the effective lifetime of a DNS system is a significant source of uncertainty in the merger rate contribution of each DNS system. Fortunately, the effect of the variation in the radio lifetime is seen only in pulsars of the type of J0737--3039B and J1906+0746, i.e. the second-born, non-recycled younger constituents of the DNS systems. The recycled pulsars in DNS systems have radio lifetimes longer than the coalescence time by emission of gravitational waves. In the Double Pulsar system, since both NSs have been detected as pulsars, we can ignore pulsar B in that system. However, the companion neutron star in the J1906+0746 system has not yet been detected as a pulsar, and we have to account for the uncertainty in the radio lifetime of the detected pulsar.

        \subsubsection{Extrapolation to LIGO's observable volume}
            
            In extrapolating from the Milky Way merger rate to the merger detection rate, we assumed that the DNS merger rate is accurately traced by the massive star formation rate in galaxies, which in turn can be traced by the B-band luminosity of the galaxies. This assumption might lead to an underestimation of the contribution of elliptical and dwarf galaxies to the merger detection rate for LIGO. As an example, the lack of current star formation in elliptical galaxies implies that binaries of the J1757--1854, J1946+2052 and J0737--3039 type might have already merged. However, there might be a population of DNS systems like B1534+12 and J1756--2251 in those galaxies which are due for mergers around the current epoch. However, as we see in this analysis, systems such as B1534+12 and J1756--2251 are not large contributors to the Galactic merger rate, and should not drastically affect the merger detection rate. 
            
            The GW170817 DNS merger event was localized to an early type host galaxy \citep[][]{THE_DNS_merger_loc}, NGC 4993. \citet[][]{THE_DNS_host_gal_prop} concluded that NGC 4993 is a normal elliptical galaxy, with an SB profile consistent with a bulge-dominated galaxy. However, this galaxy shows evidence for having undergone a recent merger event \citep[][]{THE_DNS_host_gal_prop}, which might have triggered star formation in the galaxy. Thus, the GW170817 merger cannot conclusively establish the presence of a significant number of DNS mergers in elliptical galaxies. NGC 4993 is also included in the catalog published by \citet[][]{extrapolate_to_get_LIGO_rate}, with a $B$-band luminosity of $L_{B} = 1.69 \times 10^{10} L_{B, \odot}$ and contributes in the derivation of Eq.~\ref{extrapol} \citep{extrapolate_to_get_LIGO_rate}.
            
            \citet[][]{extrapolate_to_get_LIGO_rate} estimate that the correction to the merger detection rate from the inclusion of elliptical galaxies should not be more than a factor of 1.5. Folding this constant factor into our calculation, our merger detection rate for LIGO increases to $0.27^{+0.20}_{-0.09}$~yr$^{-1}$.
            
        \subsubsection{Unobserved underlying DNS population in the Milky Way}
            
            In this analysis, we assume that the population of the DNS systems that has been detected accurately represents the ``true" distribution of the DNS systems in the Milky Way. It is possible that there exists a population of DNS systems which has been impossible to detect due to a combination of small fluxes from the pulsar in the system, extreme Doppler smearing of the orbit (for relativistic systems such as J0737--3039) and extremely large beaming correction factors (i.e. very narrow beams). Addition of more DNS systems, particularly highly relativistic systems with large beaming correction factors, would lead to an increase in the Milky Way merger rate, which would consequently lead to an increase in the merger detection rate for LIGO.
            
    \subsection{Comparison with other DNS merger rate estimates}
        
        \begin{figure*}[htb]
            \centering
            \includegraphics[width = \textwidth]{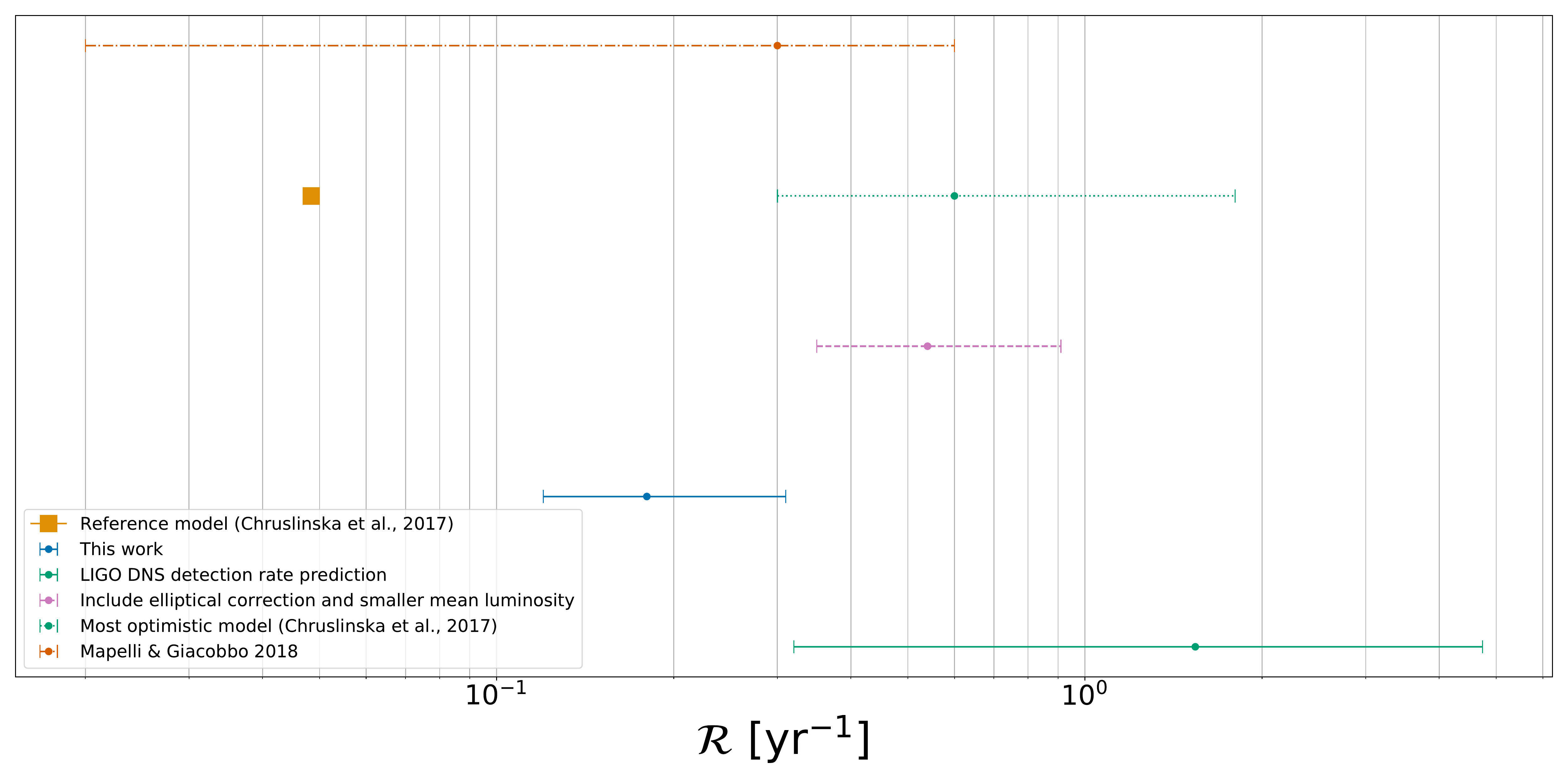}
            \caption{We compare the merger detection rate derived in this work with those derived in works by \citet[][]{chruslinska_rate} and \citet[][]{mapelli_rate}. We also plot our estimate of the merger detection rate including the correction for elliptical galaxies and a lower luminosity population of DNS systems in the Galaxy. We can see that the rate derived in \citet[][]{chruslinska_rate} with their reference model is significantly lower than that predicted in this work, while their most optimistic model is consistent with our results at 90\% confidence. The merger detection rates predicted by \citet[][]{mapelli_rate} are consistent with those derived in this work.}
            \label{compare_rates}
        \end{figure*}
        
        We can also compare our merger detection rate to that predicted through theoretical studies and simulations of the formation and evolution of DNS binary systems. This approach to calculating the merger detection rate factors in the different evolutionary scenarios leading to the formation of the DNS system, including modelling stellar wind in progenitor massive star binaries, core-collapse and electron-capture supernovae explosions, natal kicks to the NSs and the common-envelope phase \citep[][]{abadie_dns_merger_rate, dominik_dns_merger_rate_1, dominik_dns_merger_rate_2}. We compare our merger detection rate to the predictions made using the above methodology following the DNS merger detected by LIGO \citep{THE_DNS_merger}, i.e. the studies by \citet[][]{chruslinska_rate} and \citet[][]{mapelli_rate}. We plot their estimates along with those calculated in this work in Fig.~\ref{compare_rates}.
        
        \citet[][]{chruslinska_rate}, using their reference model, calculated a merger detection rate density for LIGO of 48.4~Gpc$^{-3}$~yr$^{-1}$, which scaled to a range distance of 100~Mpc is equivalent to a merger detection rate of 0.0484~yr$^{-1}$. This value is significantly lower than our range of predicted merger detection rates. In addition to the reference model, they also calculate the merger detection rate densities for a variety of different models, with the most optimistic model predicting a merger detection rate density of $600^{+600}_{-300}$~Gpc$^{-3}$~yr$^{-1}$. Scaling this to our reference range distance of 100~Mpc, we obtain a merger detection rate of $0.6^{+0.6}_{-0.3}$~yr$^{-1}$, which is consistent with the LIGO calculated merger detection rate ($\mathcal{R}_{\rm A17} = 1.54^{+3.20}_{-1.22}$~yr$^{-1}$). However, this optimistic model assumes that Hertzsprung gap (HG) donors avoid merging with their binary companions during the common-envelope phase. Applying the same evolutionary scenario to black hole binaries (BHBs) overestimates their merger detection rate from that derived using the BHB mergers observed by LIGO \citep{chruslinska_rate}. Thus, for the optimistic model to be correct would need the common-envelope process to work differently for BHB systems as compared to DNS systems, or that BHB systems would endure a bigger natal kick in the same formation scenario than DNS systems would \citep{chruslinska_rate}.
        
        \citet[][]{mapelli_rate} showed that the above problem could be avoided and a rate consistent with the LIGO prediction of the merger detection rate ($\mathcal{R}_{\rm A17} = 1.54^{+3.20}_{-1.22}$~yr$^{-1}$) could be obtained if there is high efficiency in the energy transfer during the common-envelope phase coupled with low kicks for both electron capture and core-collapse supernovae. Based on their population synthesis, they calculate a merger detection rate density of $\sim 600$~Gpc$^{-3}$~yr$^{-1}$ \citep[for $\alpha = 5$, low $\sigma$ in Fig.~1][]{mapelli_rate}. The full range of merger detection rate densities predicted by \citet[][]{mapelli_rate} ranges from $\sim 20$~Gpc$^{-3}$~yr$^{-1}$ to $\sim 600$~Gpc$^{-3}$~yr$^{-1}$, which at a range distance of 100~Mpc corresponds to a merger detection rate ranging from $0.02$~yr$^{-1}$ to $0.6$~yr$^{-1}$. This merger detection rate is consistent with that derived in this work, and lends credence to the hypotheses of high energy transfer efficiency in the common-envelope phase and low natal kicks in DNS systems made by \citet[][]{mapelli_rate}.
        
    \subsection{Future prospects}
        
        In the short term, the difference between our merger rate and that calculated by LIGO can be clarified from the results of the third operating run (O3), which is scheduled to run sometime in early 2019. Based on the fiducial model in our analysis and the predicted range distance of 120--170~Mpc for O3 \citep{LIGO_horizon_dist}, we predict, accounting for 90\% confidence intervals, that LIGO--Virgo will detect anywhere between zero and two DNS mergers. Further detections or non-detections by LIGO will be able to shed light on the detection rate within LIGO's observable volume. In addition, the localization of these mergers to their host galaxies as demonstrated by GW170817 \citep[][]{LIGO_localization} will determine the contribution of galaxies lacking in blue luminosity (such as ellipticals) to the total merger rate.
        
        In the long term, with the advent of new large scale telescope facilities such as the Square Kilometer Array \citep{SKA}, we should be able to survey our Galaxy with a much higher sensitivity. Such deep surveys might reveal more of the DNS population in our Galaxy, which would yield a better constraint on the Galactic merger rate.
        
        In addition to future radio surveys, a large number of LIGO detections of DNS mergers will allow us to probe the underlying DNS population directly. Assuming no large deviations from the DNS population parameters adopted in this study (see Sec.~\ref{survey_sims} and Sec.~\ref{discuss}), a significantly larger number of DNS merger detections by LIGO would imply a larger underlying DNS population. The localization of the DNS mergers to their host galaxies will allow us to test the variation in the DNS population with respect to host galaxy morphology. We might also be able to test if the DNS population in different galaxies is similar to the DNS population in the Milky Way. This will clarify the effect of the host galaxy morphology on the evolutionary scenario of DNS systems. 
        
\acknowledgements

MAM, DRL, and NP are members of the NANOGrav Physics Frontiers Center (NSF PHY-1430284). M.A.M and N.P. are supported by NSF AAG-1517003.  MAM and DRL have additional support from NSF OIA-1458952 and DRL acknowledges support from the Research Corporation for Scientific Advancement and NSF AAG-1616042.
%
    
\bibliography{bibliography.bib}
\bibliographystyle{aasjournal}

\end{document}